\documentclass[english]{sobraep}
\usepackage{xurl}
\usepackage{graphicx}
\usepackage[numbers]{natbib}
\usepackage{afterpage,lipsum}
\usepackage[useregional]{datetime2}
\usepackage{units}

\title{SKATEBOARDING STANCE AND HANDEDNESS: A BRIEF ANALYSIS OF RELATIONSHIP, PROPORTIONS AND INFLUENCES}

\author{Mascarenhas Alexandre\\
	\normalsize University of Tukuba, Tsukuba -- Ibaraki, Japan\\
	\normalsize website: \url{mascarenhasav.github.io}\\
	\today
}

\begin{document}

\maketitle

\begin{abstract}
	The aim of this paper is to show the relationship that lies in the fact of a person being right or left handed, in their skateboarding stance. Starting from the null hypothesis that there is no relationship, the Pearson's $\chi^{2}$ with Yates correction tests, as well as its respective p-value will be used to test the hypothesis. It will also be calculated and analyzed the residuals, Cramer's V and the Risk and Odds Ratios, with their respective confidence intervals to know the intensity of the association.
\end{abstract}

\begin{keywords}
	Skateboarding stance, Handedness, $\chi^{2}$ test, Cramer's V test, Risk and Odds Ratio
\end{keywords}





\section{INTRODUCTION}

The preference for one side of the body to carry out the most diverse tasks is not exclusive to human beings \citeonline{lateralized}. However, even in these, the origin of why a preference exists is not agreed upon to this day \citeonline{Vallortigara2005survivalWA}. There are many studies that suggest the number of people who prefer the right or left side, but defining the ideal parameters for this research is not easy. It is estimated that when strong restrictions are placed in the research, the number of left-handed people is something around 9.3\% of the population and when we consider weaker restrictions, this number can reach 18.1\% \citeonline{meta-analysis}.

There is some evidence to suggest that there is no strong relationship between tasks that are unilateral and those that are bilateral, for example, throwing and hitting a ball \citeonline{baseball}. In addition, when related to sports, some authors make due warnings, (i) the term "handness" should be used with more caution in the context of sports-related laterality research and (ii) the observation of lateral preferences in sports may be of little aptitude for the verification of evolutionary theories of the hand \citeonline{sport-tasks}.

Knowledge of the origin of handedness and its relationship with certain tasks can be of great help in understanding how the laterality of the brain works, and how it may be related to other human abilities. This would also help in the design and production of various products that take into account the handedness. The results obtained from this research are intended to provide further data for this understanding.

When related specifically to sports, the considerations that need to be taken into account are a little different with regard to the handedness of the person. This is because, in general, the handedness of the population being evaluated is defined by means of activities which have little or almost no relation to the specific task which the sport requires \citeonline{sport-tasks}.

\section{ORGANIZATION OF THE PAPER}

This paper will be organized like follows:

\begin{itemize}
    \item \textbf{Aim of the experiment:} it will be briefly discussed what stance is in skateboarding as well as its relevance in the world of the sport which justifies the intention of the experiment;
	\item \textbf{How was the data obtained:} it will be described how the choice of the sample size of the experiment was made, a task that, because it deals with qualitative variables, there is a long discussion about. As well as which was the method used to obtain the data to be analyzed;
	\item \textbf{Pre-processing of the data:} it is common that a dataset has some issues in its data, such as data missing and outliers. Will be described which issues were found in the datasets and how they were treated;
	\item \textbf{Analysis and presentation of the data:} in the data analysis, all statistics used for the null hypothesis test will be described, and their values and confidence intervals will be shown by means of graphs and tables;
	\item \textbf{Conclusion:} the conclusion based on the results obtained in the data analysis, and the possible inferences that can be made based on them. As well as possible future studies and other areas of knowledge in which these results are inserted.
\end{itemize}

\section{EXPERIMENT}

\subsection{Scenario}

\begin{figure}[t!]
  \centering
  \includegraphics[width=0.8\columnwidth]{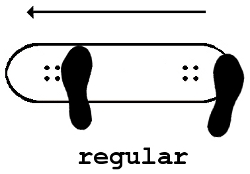}
  \caption{A representation of the Regular stance \citeonline{streetwar}}
  \label{fig:fig1}
\end{figure}

\begin{figure}[t!]
  \centering
  \includegraphics[width=0.8\columnwidth]{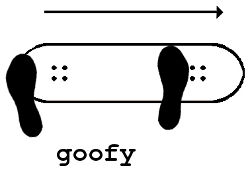}
  \caption{A representation of the Goofy stance \citeonline{streetwar}}
  \label{fig:fig2}
\end{figure}

On the skateboard, an important characteristic is stance\footnote{From here on, only the term stance will be used to refer to skateboard stance}, which is the preferred position in which a person normally stands when climbs on a skateboard. The two possible stances are described as follows and can be seen in Figure \ref{fig:fig1} and \ref{fig:fig2}: 
\begin{itemize}
    \item \textbf{Regular:} When the skater's right foot is on the back of the skateboard\footnote{The back of the skateboard is called the tail and the front is called the nose.}; and
    \item \textbf{Goofy:} When the skater's left foot is on the back of the skateboard.
\end{itemize}

In Skateboarding competitions, it is important to know the stance of each skateboarder, because when performing a trick in the stance that is not the preferred one, this becomes worth more points, that's because at first, it is more difficult to perform it this way. When the trick is done like this, it is said to be done by \textit{switch}.

Thus, the aim of this experiment is, by means of real data collected in the field, to carry out some statistical tests and to verify if there is any preference of skateboarding stance with handedness.

\subsection{Obtaining the data}

The idea is to get individual data on each person's handedness and stance. Since these are qualitative variables, the calculation of the sample size is the subject of much discussion and is not straightforward. The method and how the sample size was defined will be described as follows. 

\subsubsection{Method}

The method used to obtain data on handedness and skateboarding stance was the individual interviews per person. In interview-based surveys the quality of the questions is a very important factor \citeonline{interview}. Thus, to avoid the risk of misinterpretation of the questions asked, most of the interviewees were questioned in person\footnote{By the author of this paper together with contributors.} (and not by internet, email, etc.), and with the presence of a skateboard so that there would be no doubt as to the person's stance. 

Therefore,the questions on the interview form consisted of two mandatory and two optional questions, as follows:

\begin{itemize}
    \item \textbf{Handedness (mandatory):} The person's handedness was initially asked. If there was doubt, or contradictions between the preference of using the hands or the feet, the preference was to use the feet in activities such as kicking a ball. The two possible answers were: \textit{Right} or \textit{Left}.
    
    \item \textbf{Stance (mandatory):} Initially it was asked which foot the person would place on the back of the skateboard, considering a direction of motion. If in doubt, the person was asked to get on the skateboard with the help of the interviewers to make sure. The two possible answers were: \textit{Regular} or \textit{Goofy}.
    
    \item \textbf{Gender (optional):} This is a question that eventually caused some doubt with sexual orientation or preference. But at first, the two pre-defined answers were: \textit{Male} and \textit{Female}. However, a text field was left in case the person interviewed had another answer.
    
    \item \textbf{Skateboarder or not (optional):} This question aimed to find out whether or not the person considered him/herself a skateboarder. However, for the purposes of the survey, the most valuable information was whether the person had ever skateboarded before.
\end{itemize}

\subsubsection{Sample size}

The composition and size of the sample are crucial parameters in the analysis of data in qualitative experiments. The quality and reliability of qualitative research requires important considerations \citeonline{sample-size2}.  There are various discussions about the best way to calculate sample size in qualitative research. Unlike quantitative surveys, where there are practical, intuitive, and very accurate methods to arrive at the required sample size values \citeonline{sample-size3}.

In \citeonline{sample-size1} it is possible to see an analysis of a large number of articles (214) on qualitative research, where the various justifications used by researchers for the sample size used are elucidated. Where the most commonly\footnote{55.4\% of all justifications invoked by the researchers was that they achieved the \textit{data saturation} \citeonline{sample-size1}} principle used was the \textit{saturation}, which is characterized by the moment when a larger amount of data no longer provides new information.
However, some researches in the field shows that for research related to medicine and psychology, a number between 20-40 interviewed is usually sufficient to demonstrate the expected effects for qualitative research \citeonline{sample-size1}.

For this experiment, a mixed approach between the numbers proposed by researchers and the concept of saturation was adopted. Initially, 33 interviewees' data were acquired, and a brief pre-analysis of the data was made to get an idea of the results, and then 5 more cycles of acquisition of the same number of data. This way, it was possible to verify when the saturation concept was becoming apparent in the samples. In all, 165 people were interviewed.

\subsection{Pre-processing}
Before effectively starting the data analysis, it was made a pre-processing of the data. This is important, because allows to identify some issues with the dataset, such as, missing data, input error, incompatibility. The interview form provided the interviewees' answers directly into a spreadsheet in csv (comma-separated values\footnote{A comma-separated values file is a delimited text file that uses a comma to separate values. Each line of the file is a data record. Each record consists of one or more fields, separated by commas\citeonline{wiki}}) format. To make it easier for the users to answer, the alternatives were written in full (e.g. Regular). However, for data analysis, especially in the calculation of statistical tests, it is more convenient to work with binary variables (such as 0 and 1, R and G, etc.). Thus, a pre-processing of the data was made in order to make the dataset suitable for statistical calculations. After this, The data was organized, as can be seen in Table \ref{table:values}.

\begin{table}[t!]
    \hrulefill
	\centering
	\caption{\textbf{Number of people classified according to their Handedness and Stance}}
	\footnotesize
	\setlength{\tabcolsep}{16pt}
	\begin{tabular}{cccc}
		\toprule [1pt]	
		\hline
		\multirow{2}{*}{\textbf{\nicefrac{Handedness }{ Stance}}} & 
		\multirow{2}{*}{\centering \textbf{Regular}} &
		\multirow{2}{*}{\centering \textbf{Goofy}} & 
		\multirow{2}{*}{\centering \textbf{Total}} \cr \\ 	
		\hline
		\multirow{2}{*}{Right-handed} & 
		\multirow{2}{*}{\centering 65} & 
		\multirow{2}{*}{\centering 75} & 
		\multirow{2}{*}{\centering 140} \cr \\ 
		\hline
		\multirow{2}{*}{Left-handed} & 
		\multirow{2}{*}{\centering 13} & 
		\multirow{2}{*}{\centering 12} & 
		\multirow{2}{*}{\centering 25} \cr \\ 
		\hline
		\multirow{2}{*}{Total} & 
		\multirow{2}{*}{\centering 78} & 
		\multirow{2}{*}{\centering 87} & 
		\multirow{2}{*}{\centering 165} \cr \\ 
		\hline
		\bottomrule[1pt]
	\end{tabular} \label{table:values}
    \end{table}

\begin{figure}[h]
  \centering
  \includegraphics[width=\columnwidth]{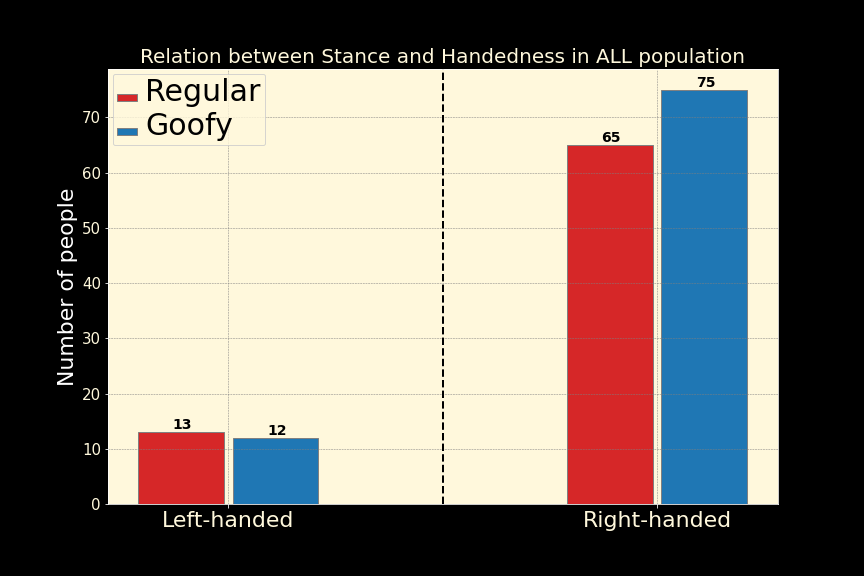}
  \caption{Bar graph representing the proportions of the stances for each handedness of all data}
  \label{fig:bar_all}
\end{figure}

\begin{figure}[h]
  \centering
  \includegraphics[width=\columnwidth]{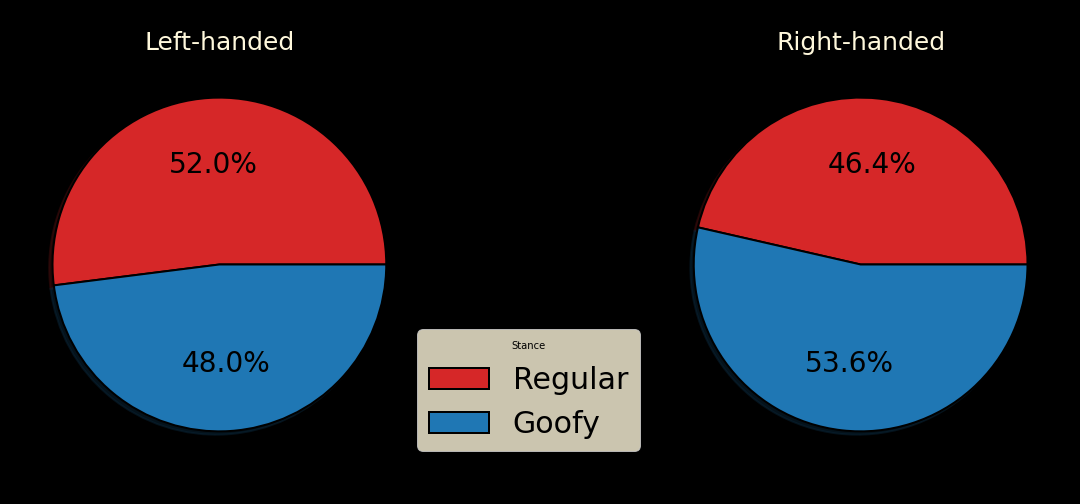}
  \caption{Pie chart representing the proportions of the stances for each handedness of all data}
  \label{fig:pie_all}
\end{figure}

    With this data, it is possible to apply the statistical test calculations and characterize whether or not there is a relationship between the Handedness and Stance variables, and if so, how strong it is. But before that, in Figures \ref{fig:bar_all} and \ref{fig:pie_all}, it is shows a bar graph and a pie chart of the data, respectively, where it is possible visually see that if there is a relationship between the variables, in the data from this experiment it was not pronounced.

\subsection{Analysis and presentation}
As a first analysis, we will start from the null hypothesis that there is no relationship between a person's Handedness and Stance. The first statistical test will be Pearson's $\chi^{2}$ squared, one of the most widely used tests to measure association between two categorical variables \citeonline{chi2}. As the $\chi^{2}$ test does not at first provide the strength of correlation, the residuals, Cramer's V coefficient, Risk and Odds Ratios will be calculated next.

 \begin{table}[t!]
	\centering
    \hrulefill
	\caption{\textbf{Number of people expected, considering no relationship between Handedness and Stance}}
	\footnotesize
	\setlength{\tabcolsep}{16pt}
	\begin{tabular}{cccc}
		\toprule [1pt]	
		\hline
		\multirow{2}{*}{\textbf{\nicefrac{Handedness }{ Stance}}} & 
		\multirow{2}{*}{\centering \textbf{Regular}} &
		\multirow{2}{*}{\centering \textbf{Goofy}} & 
		\multirow{2}{*}{\centering \textbf{Total}} \cr \\	
		\hline
		\multirow{2}{*}{Right-handed} & 
		\multirow{2}{*}{\centering 66.19} & 
		\multirow{2}{*}{\centering 73.82} & 
		\multirow{2}{*}{\centering 140} \cr \\
		\hline
		\multirow{2}{*}{Left-handed} & 
		\multirow{2}{*}{\centering 11.82} & 
		\multirow{2}{*}{\centering 13.18} & 
		\multirow{2}{*}{\centering 25} \cr \\
		\hline
		\multirow{2}{*}{Total} & 
		\multirow{2}{*}{\centering 78} & 
		\multirow{2}{*}{\centering 87} & 
		\multirow{2}{*}{\centering 165} \cr \\
		\hline
		\bottomrule[1pt]
	\end{tabular} \label{table:expected}
    \end{table}

\subsubsection{Pearson's $\chi^{2}$ test}
The first step is to calculate the expected frequency for each cell in Table \ref{table:values}. This gives the expected number of people in each classification, considering there is no relationship between Handedness and Stance. The result can be seen in Table \ref{table:expected}. In using the $\chi^{2}$ distribution in the test of association, a continuous probability distribution is being used to approximate discrete probabilities. A correction, attributable to Yates, can be applied to the frequencies to make the test closer to the exact test \citeonline{review8}. Its recommended by \citeonline{bland} that for a small number of data, the correction be made by the Yate's method. This basically consists of adding 0.5 to the smallest value found in the expected frequency table, and adjusting all the others to fit within the total values. The expected frequency values are now those of Table \ref{table:yates} and these will be used for the next calculations.

 \begin{table}[h!]
	\centering
    \hrulefill
	\caption{\textbf{Number of people expected, adjusted by Yates' correction.}}
	\footnotesize
	\setlength{\tabcolsep}{16pt}
	\begin{tabular}{cccc}
		\toprule [1pt]	
		\hline
		\multirow{2}{*}{\textbf{\nicefrac{Handedness }{ Stance}}} & 
		\multirow{2}{*}{\centering \textbf{Regular}} &
		\multirow{2}{*}{\centering \textbf{Goofy}} & 
		\multirow{2}{*}{\centering \textbf{Total}} \cr \\	
		\hline
		\multirow{2}{*}{Right-handed} & 
		\multirow{2}{*}{\centering 65.68} & 
		\multirow{2}{*}{\centering 74.32} & 
		\multirow{2}{*}{\centering 140} \cr \\
		\hline
		\multirow{2}{*}{Left-handed} & 
		\multirow{2}{*}{\centering 12.32} & 
		\multirow{2}{*}{\centering 12.68} & 
		\multirow{2}{*}{\centering 25} \cr \\
		\hline
		\multirow{2}{*}{Total} & 
		\multirow{2}{*}{\centering 78} & 
		\multirow{2}{*}{\centering 87} & 
		\multirow{2}{*}{\centering 164} \cr \\
		\hline
		\bottomrule[1pt]
	\end{tabular} \label{table:yates}
\end{table}

The $\chi^{2}$ calculation involves the sum of the differences between the expected frequencies (Table \ref{table:yates}) of each cell and the actual observed value (Table \ref{table:values}), and for a r rows and c columns table it is given by: 
\begin{equation}
    \sum_{i=1}^{r}\sum_{j=1}^{c}\frac{(O_{ij} - E_{ij})^{2}}{E_{ij}}
\end{equation}
Where $O_{ij}$ is the observed frequency and $E_{ij}$ is the expected frequency in the cell in row i and column j.
For the experimental data, the statistical test value found was $\chi^{2} = 0.09$ with 1 degree of freedom\footnote{degrees of freedom is founded with (r – 1)(c - 1), where r is the number of rows and c the number of columns}, which leads to a \textit{p-value}$= 0.77$. This value strongly suggests that the null hypothesis is not rejected, consequently showing that, in principle, there is no relationship between the variables, since a p-value of 0.77 indicates that there is a chance that approximately 77\% of the time, frequencies like these will be observed if there is no relationship between the variables.

\subsubsection{Residuals}

Since $\chi^{2}$ does not provide the strength of the association (or non association) between the variables, the residuals will be calculated. This is roughly the difference between the expected frequency values of each cell and the observed values, given by:

\begin{equation}
    d_{ij} = \frac{O_{ij} - E_{ij}}{\sqrt{E_{ij}(1-\frac{n_i}{N})(1-\frac{n_j}{N})}}
\end{equation}

where $d_{ij}$ is the adjusted standardized residual, $n_i$ is the total frequency for row i, $n_j$ is the total frequency for column j and N is total overall total frequency. Thus, the greater the difference, and consequently the residual, the more significant is the association between the variables. On the other hand, low residuals demonstrate a weak relationship between the variables. The values found are shown in Table \ref{table:residuals}.

\begin{table}[t!]
	\centering
    \hrulefill
	\caption{\textbf{The adjusted standardized residuals}}
	\footnotesize
	\setlength{\tabcolsep}{25pt}
	\begin{tabular}{ccc}
		\toprule [1pt]	
		\hline
		\multirow{2}{*}{\textbf{\nicefrac{Handedness }{ Stance}}} & 
		\multirow{2}{*}{\centering \textbf{Regular}} & 
		\multirow{2}{*}{\centering \textbf{Goofy}} \cr \\
		\hline
		\multirow{2}{*}{Right-handed} & 
		\multirow{2}{*}{\centering $-0.298$} & 
		\multirow{2}{*}{\centering $0.296$} \cr \\
		\hline
		\multirow{2}{*}{Left-handed} & 
		\multirow{2}{*}{\centering $0.290$} & 
		\multirow{2}{*}{\centering $-0.302$} \cr  \\
		\hline
		\bottomrule[1pt]
	\end{tabular} \label{table:residuals}
    \end{table}
    
Since the residuals follow a normal distribution with center at 0 and standard deviation of 1, absolute values greater than 2 are considered significant \citeonline{review2}. That is, the values in Table \ref{table:residuals} show a weak relationship between the variables.

\subsubsection{Cramer's V coefficient} This is an effect size measurement for the $\chi^{2}$ test of independence. It measures how strongly two categorical fields are associated \citeonline{IBM}. Its value is given by:

\begin{equation}
    V = \sqrt{ \frac{\frac{\chi^{2}}{N}}{min(r-1, c-1)} }
\end{equation}

And it was found a value of $V=0.02$. This is a value very close to zero, which consulting Table \ref{table:cramers} shows that the relationship between the variables is weak \citeonline{IBM}.


\subsubsection{Risk and Odds Ratio}

Another way to approach the association between two variables is to examine the consequence is their possible risk factor. This approach is generally used in the medical field, where one wants to assess the risk of developing a disease given exposure to a risk factor \citeonline{review8}. A parallel will be made, where what would be the risks of a person being, for example, Regular, given that he is right-handed, and then a comparison will be made with people who are not right-handed but are also Regular. 

From Table \ref{table:values} the risk of people who are Right-handed and Regular, considering the total number of Right-handed people, is $65/140 = 0.46$ and the risk of the person being regular being left-handed, considering the total number of left-handed people is $13/25 = 0.52$

The Risk Ratio (RR) measures the increased risk is that a person is Regular being Right-handed than one who is Left-handed, and can be calculated in the following simple way:
\begin{equation}
    RR = \frac{\text{risk for Regular Right-handers}}{\text{risk for Regular Left-handers}}
\end{equation}

Resulting in $RR = 0.89$ with 95\% confidence interval of 0.59 to 1.35. This means that a Right-handed person is 11\% less likely to be Regular than a Left-handed one.

The Odds Ratio (OR) is another way of measuring the comparison between people exposed and unexposed to a risk factor. Odds can be interpreted as the probability of something happening, given the likelihood of it not happening \citeonline{odds}. From Table \ref{table:values}, the odds of a Right-handed being a Regular by non-Right-handed Regulars is $65/13 = 5.00$ and the odds of a Right-handed being Goofy by the non right-handed Goofys is $75/12 = 6.25$. So to calculate the Odds Ratio we use the following equation:

\begin{equation}
    OR = \frac{\text{odds for Regular Right-handers}}{\text{odds for Goofy Right-handers}}
\end{equation}

Resulting in $OR = 0.80$ with 95\% confidence interval of 0.34 to 1.88. This value can be interpreted as follows \citeonline{oddsratio}, for every one (1.0) chance that a person is Regular because he is not Right-handed, there is a 0.80 chance that this person is Regular because it is Right-handed.

For both RR and OR, the possible values range from 0 to infinite, values close to 1 indicate little association between the variables, and the exact value of 1 indicates no association. For values greater than 1, the probability of such an event occurring increases, and for values less than 1, it decreases. 
It is important to note that when the confidence interval for both RR and OR passes through the value of 1, it is considered that the event is probably null in the population.

\subsubsection{Summarize}
These were the statistical tests used to characterize the association between a person's Handedness and Skateboarding Stance and Table \ref{table:summarize} summarizes the values for the tests as well as their CI interval where appropriate.

In the next subsection further consideration will be given to the data and the same tests will be performed.

\begin{table}[t!]
	\centering
    \hrulefill
	\caption{\textbf{Interpretation of effect size (Cramer's V)}}
	\footnotesize
	\setlength{\tabcolsep}{10pt}
	\begin{tabular}{cc}
		\toprule [1pt]	
		\hline
		\multirow{3}{*}{\textbf{Effect size (Cramer's V)}} & 
		\multirow{3}{*}{\centering \textbf{Interpretation}} \cr  \\ \\
		\hline
		\multirow{3}{*}{$V \leq 0.2$} & 
		\multirow{3}{4.2cm}{\centering  The result is weak. Although the result is statistically significant, the fields are only weakly associated.} \cr \\ \\
		\hline
		\multirow{2}{*}{$0.2 < V \leq 0.6$} & 
		\multirow{2}{4.2cm}{\centering The result is moderate} \cr \\
		\hline
		\multirow{2}{*}{V > 0.6} & 
		\multirow{2}{4.2cm}{\centering The result is strong. The fields are strongly associated.} \cr \\
		\hline
		\bottomrule[1pt]
	\end{tabular} \label{table:cramers}
    \end{table}

 \begin{table}[h!]
    	\centering
        \hrulefill
    	\caption{\textbf{Summary of the statistics and their CI}}
    	\footnotesize
    	\setlength{\tabcolsep}{9.2pt}
        	\begin{tabular}{ccccc}
        		\toprule [1pt]	
        		\hline
        		\multirow{2}{*}{\centering \textbf{$\chi^{2}$}} & 
        		\multirow{2}{*}{\centering \textbf{p-value}} &
        		\multirow{2}{*}{\centering \textbf{Cramer's V}} & 
        		\multirow{2}{*}{\centering \textbf{Risk Ratio}} &
        		\multirow{2}{*}{\centering \textbf{Odds Ratio}} \cr \\		
        		\hline
        		\multirow{2}{*}{\centering $0.09$} & 
        		\multirow{2}{*}{\centering $0.77$} & 
        		\multirow{2}{*}{\centering $0.02$} & 
        		\multirow{2}{*}{\centering $0.89$} & 
        		\multirow{2}{*}{\centering $0.80$} \cr \\
        		\hline
        		\multirow{2}{*}{\centering []} & 
        		\multirow{2}{*}{\centering []} & 
        		\multirow{2}{*}{\centering []} & 
        		\multirow{2}{*}{\centering [0.59, 1.35]} & 
        		\multirow{2}{*}{\centering [0.34, 1.88]} \cr \\
        		\hline
        		\bottomrule[1pt]
        	\end{tabular} \label{table:summarize}
        \end{table}

\subsection{Other considerations}

The consideration to be taken into account in the next calculations is the gender of the people interviewed. The same graphs will be shown and the same statistical tests performed as in the previous subsections. 

Before showing the data, it is important to note that because gender is a difficult question to approach, there may be several influences on the answers, and not all participants answered, or even do not identify with the options that will be analyzed here, namely, male and female. 

\subsubsection{Females}

\begin{figure}[h!]
      \centering
      \includegraphics[width=\columnwidth]{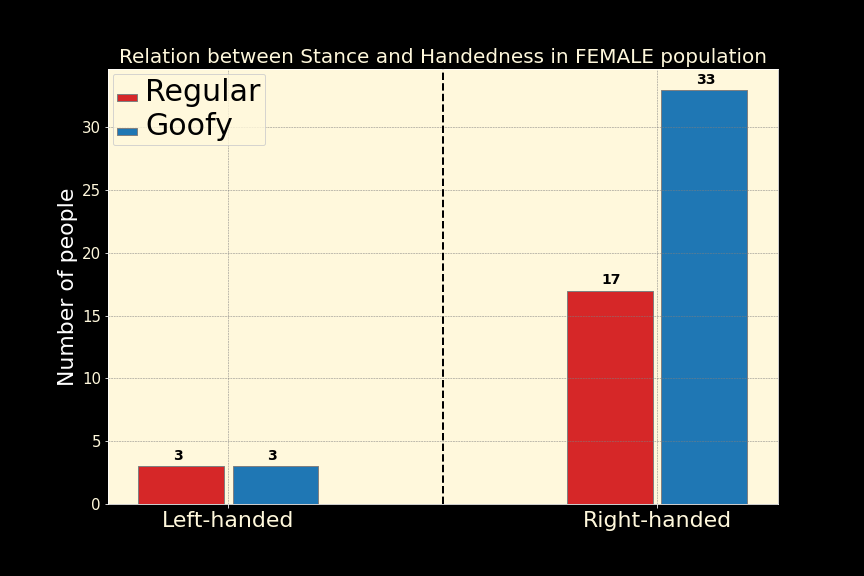}
      \caption{Bar graph representing the proportions of the base stances for each handedness by the people who identified themselves as female}
      \label{fig:bar_female}
    \end{figure}

\begin{figure}[h!]
      \centering
      \includegraphics[width=\columnwidth]{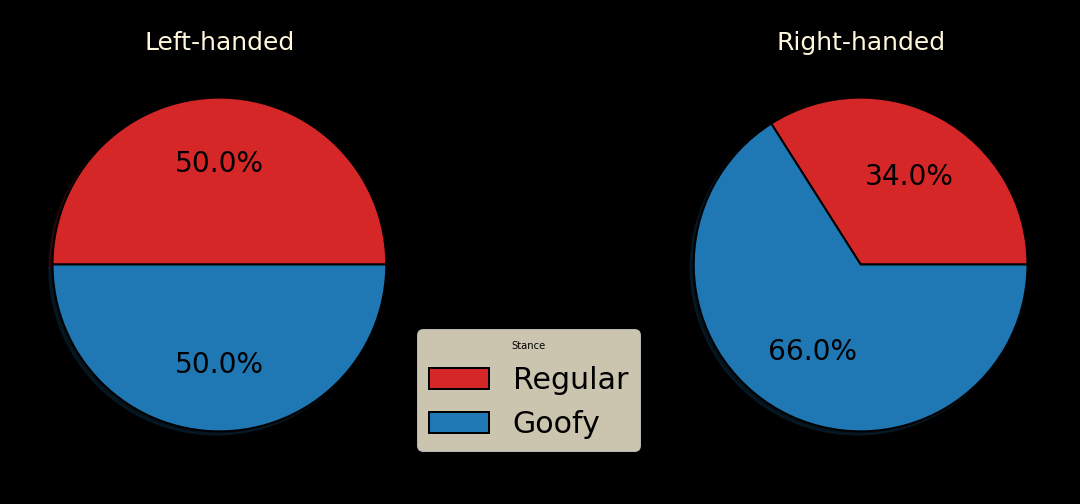}
      \caption{Pie chart representing the proportions of the base stances for each handedness by the people who identified themselves as female}
      \label{fig:pie_female}
    \end{figure}

For those who identified themselves as Female, Table \ref{table:values_F} shows the numbers referring to Stance and Handedness and Figures \ref{fig:bar_female} and \ref{fig:pie_female} are graphs showing these relationships.

    \begin{table}[h!]
    	\centering
        \hrulefill
    	\caption{\textbf{People classified according to their Handedness and Stance who identified themselves as female}}
    	\footnotesize
    	\setlength{\tabcolsep}{17pt}
    	\begin{tabular}{cccc}
    		\toprule [1pt]	
    		\hline
    		\multirow{2}{*}{\textbf{\nicefrac{Handedness }{ Stance}}} & 
    		\multirow{2}{*}{\centering \textbf{Regular}} &
    		\multirow{2}{*}{\centering \textbf{Goofy}} & 
    		\multirow{2}{*}{\centering \textbf{Total}} \cr \\ 	
    		\hline
    		\multirow{2}{*}{Right-handed} & 
    		\multirow{2}{*}{\centering 17} & 
    		\multirow{2}{*}{\centering 33} & 
    		\multirow{2}{*}{\centering 50} \cr \\ 
    		\hline
    		\multirow{2}{*}{Left-handed} & 
    		\multirow{2}{*}{\centering 3} & 
    		\multirow{2}{*}{\centering 3} & 
    		\multirow{2}{*}{\centering 6} \cr \\ 
    		\hline
    		\multirow{2}{*}{Total} & 
    		\multirow{2}{*}{\centering 20} & 
    		\multirow{2}{*}{\centering 36} & 
    		\multirow{2}{*}{\centering 55} \cr \\ 
    		\hline
    		\bottomrule[1pt]
    	\end{tabular} \label{table:values_F}
    \end{table}

\break
    \begin{table}[t!]
    	\centering
        \hrulefill
    	\caption{\textbf{Summary of the calculated statistics and their CI for people who identified themselves as female}}
    	\footnotesize
    	\setlength{\tabcolsep}{9.2pt}
        	\begin{tabular}{ccccc}
        \toprule [1pt]	
        \hline
        \multirow{2}{*}{\centering \textbf{$\chi^{2}$}} & 
        \multirow{2}{*}{\centering \textbf{p-value}} &
        \multirow{2}{*}{\centering \textbf{Cramer's V}} & 
        \multirow{2}{*}{\centering \textbf{Risk Ratio}} &
        \multirow{2}{*}{\centering \textbf{Odds Ratio}} \cr \\		
        \hline
        \multirow{2}{*}{\centering $0.10$} & 
        \multirow{2}{*}{\centering $0.75$} & 
        \multirow{2}{*}{\centering $0.04$} & 
        \multirow{2}{*}{\centering $0.68$} & 
        \multirow{2}{*}{\centering $0.52$} \cr \\
        \hline
        \multirow{2}{*}{\centering []} & 
        \multirow{2}{*}{\centering []} & 
        \multirow{2}{*}{\centering []} & 
        \multirow{2}{*}{\centering [0.28, 1.65]} & 
        \multirow{2}{*}{\centering [0.09, 2.83]} \cr \\
        \hline
        \bottomrule[1pt]
    \end{tabular} \label{table:TableI}
    \end{table}
    \hfill \break

\subsubsection{Males}

\begin{figure}[h!]
  \centering
  \includegraphics[width=\columnwidth]{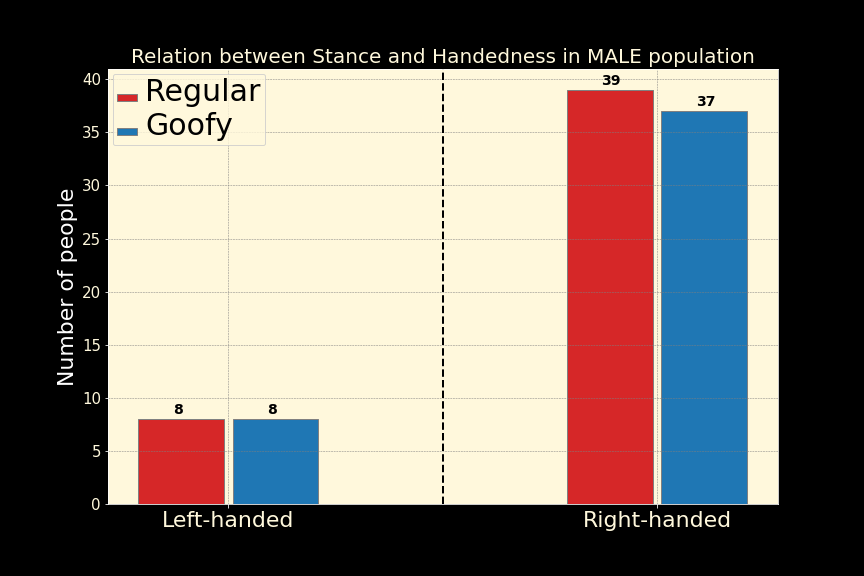}
  \caption{Bar graph representing the proportions of the base stances for each handedness by the people who identified themselves as male}
  \label{fig:bar_male}
\end{figure}

\begin{figure}[h!]
  \centering
  \includegraphics[width=\columnwidth]{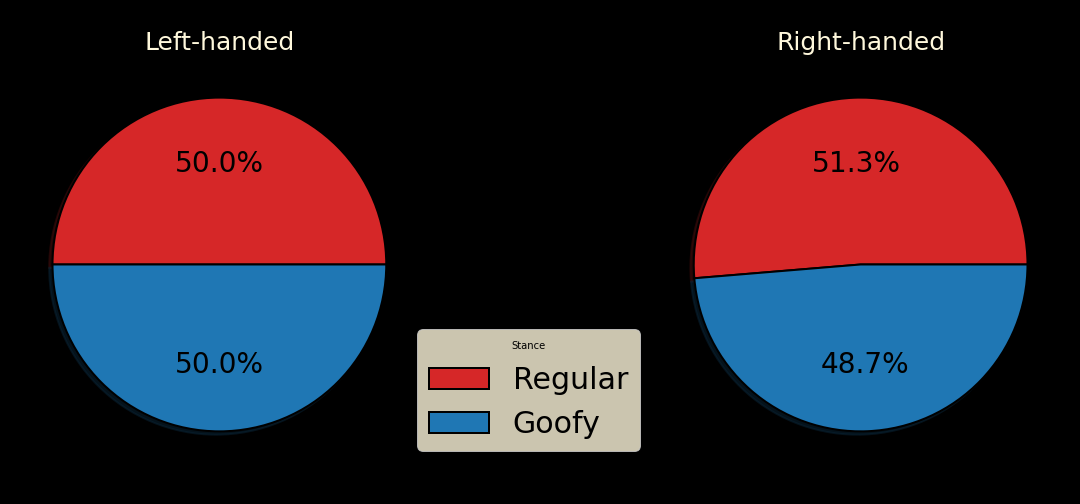}
  \caption{Pie chart representing the proportions of the base stances for each handedness by the people who identified themselves as male}
  \label{fig:pie_male}
\end{figure}

For those who identified themselves as Female, Table \ref{table:values_M} shows the numbers referring to Stance and Handedness and Figures \ref{fig:bar_male} and \ref{fig:pie_male} are graphs showing these relationships.

\begin{table}[h!]
	\centering
    \hrulefill
	\caption{\textbf{People classified according to their Handedness and Stance who identified themselves as male}}
	\footnotesize
	\setlength{\tabcolsep}{17pt}
	\begin{tabular}{cccc}
		\toprule [1pt]	
		\hline
		\multirow{2}{*}{\textbf{\nicefrac{Handedness }{ Stance}}} & 
		\multirow{2}{*}{\centering \textbf{Regular}} &
		\multirow{2}{*}{\centering \textbf{Goofy}} & 
		\multirow{2}{*}{\centering \textbf{Total}} \cr \\ 	
		\hline
		\multirow{2}{*}{Right-handed} & 
		\multirow{2}{*}{\centering 39} & 
		\multirow{2}{*}{\centering 37} & 
		\multirow{2}{*}{\centering 76} \cr \\ 
		\hline
		\multirow{2}{*}{Left-handed} & 
		\multirow{2}{*}{\centering 8} & 
		\multirow{2}{*}{\centering 8} & 
		\multirow{2}{*}{\centering 16} \cr \\ 
		\hline
		\multirow{2}{*}{Total} & 
		\multirow{2}{*}{\centering 47} & 
		\multirow{2}{*}{\centering 45} & 
		\multirow{2}{*}{\centering 92} \cr \\ 
		\hline
		\bottomrule[1pt]
	\end{tabular} \label{table:values_M}
    \end{table}

 \begin{table}[t!]
    	\centering
        \hrulefill
    	\caption{\textbf{Summary of the calculated statistics and their CI for people who identified themselves as male}}
    	\footnotesize
    	\setlength{\tabcolsep}{9.2pt}
        	\begin{tabular}{ccccc}
        		\toprule [1pt]	
        		\hline
        		\multirow{2}{*}{\centering \textbf{$\chi^{2}$}} & 
        		\multirow{2}{*}{\centering \textbf{p-value}} &
        		\multirow{2}{*}{\centering \textbf{Cramer's V}} & 
        		\multirow{2}{*}{\centering \textbf{Risk Ratio}} &
        		\multirow{2}{*}{\centering \textbf{Odds Ratio}} \cr \\		
        		\hline
        		\multirow{2}{*}{\centering $0.14$} & 
        		\multirow{2}{*}{\centering $0.71$} & 
        		\multirow{2}{*}{\centering $0.04$} & 
        		\multirow{2}{*}{\centering $1.03$} & 
        		\multirow{2}{*}{\centering $1.05$} \cr \\
        		\hline
        		\multirow{2}{*}{\centering []} & 
        		\multirow{2}{*}{\centering []} & 
        		\multirow{2}{*}{\centering []} & 
        		\multirow{2}{*}{\centering [0.60, 1.76]} & 
        		\multirow{2}{*}{\centering [0.36, 3.10]} \cr \\
        		\hline
        		\bottomrule[1pt]
        	\end{tabular} \label{table:TableI}
        \end{table}

\section{INSTRUCTIONS AND LIMITATIONS}

The dataset (csv file) containing all data collected in this experiment as well the python script used to evaluated the statistics and generate the graphs can be found in this link: \url{https://github.com/mascarenhasav/skateboard_stance}. They are free to use and reproduce.

Once with the chi-squared value, the power of the test was calculated, and the value found was approximately of 25\%. Being a low value, having a high probability (75\%) of committing a Type II error and incorrectly not rejecting the null hypothesis. One way to increase the power of the test would be to increase the total number of data collected, so for a test power of 85\% the suggested minimum value of the number of intreviewees would be 800 people.

\section{CONCLUSIONS}
This experiment started from the null hypothesis that there was no relationship between the stance of the skateboard and the handedness of a person. Considering the calculated statistics with their respective confidence intervals, the experiment suggests, at first, three main conclusions, considering all data, without gender distinction:
\begin{enumerate}

    \item That there is probably no relationship between the variables, that is, they are independent variables, suggesting that the null hypothesis may be correct. In other words, knowing whether a person is right-handed or left-handed tells us nothing about his or her skateboarding stance.
    
    \item The percentage of Regular, or Goofy, people is approximately 50\%, for both right-handed and left-handed people. This is a particularly interesting finding, as it contradicts the common sense that if a person is right-handed they will always prefer to use the same side to perform specific tasks.

    \item Apparently, in fact, there are more right-handed people than left-handed people in general.
\end{enumerate}  

When taken into account gender, some important points can be highlighted. At first, among those who identified themselves as Male, the proportions were very similar to the total population, and the calculated statistics suggest even more precisely that there is no relationship between the variables, with the RR and OR values very close to 1. For those who identified themselves as Female, the graphs show a possible tendency towards the Goofy stance for right-handed people. However, as the sample size for this group is considerably small, it is an effect with little statistical significance to conclude anything, and more data is needed to reinforce this idea.

Because it is a question specifically without much related research, and related to human behavior, perhaps a larger amount of data would be necessary for the conclusions to be more assertive. Even though the interviews were random, as it was done in only one city at first, naturally certain ethnic groups were not evaluated, having been restricted to individuals from one metropolitan region in Brazil. Thus, a possible future research in this area, would be to obtain more data, and ensure a great ethnic and gender diversity among the interviewees.

\bibliographystyle{bib_sobraep}
\bibliography{references}

\begin{thebibliography}{10}
\newcommand{\enquote}[1]{``#1''}
\providecommand{\url}[1]{{\tt #1}}
\providecommand{\urlprefix}{URL: }
\expandafter\ifx\csname urlstyle\endcsname\relax
  \providecommand{\doi}[1]{doi:\discretionary{}{}{}#1}\else
  \providecommand{\doi}{doi:\discretionary{}{}{}\begingroup
  \urlstyle{rm}\Url}\fi
\providecommand{\eprint}[2][]{\url{#2}}

\bibitem{lateralized}
A.~A. Beaton, \enquote{The lateralized brain: the neuroscience and evolution of
  hemispheric asymmetries}, {\em Behavioral and Brain Sciences\/}, 2018.

\bibitem{Vallortigara2005survivalWA}
G.~Vallortigara, L.~J. Rogers, \enquote{survival with an asymmetrical brain:
  advantages and disadvantages of cerebral lateralization}, {\em Behavioral and
  Brain Sciences\/}, vol.~28, pp. 575 -- 589, 2005.

\bibitem{meta-analysis}
M.~Papadatou-Pastou, E.~Ntolka, J.~Schmitz, M.~Martin, M.~R. Munafò,
  S.~Ocklenburg, S.~Paracchini, \enquote{Human handedness: A meta-analysis},
  {\em Psychol Bull\/}, vol.~6, no. 146, pp. 481--524, 2020.

\bibitem{baseball}
F.~M.~C. J~M~McLean, \enquote{Bimanual Dexterity in Major League Baseball
  Players: A Statistical Study}, {\em New England Journal of Medicine\/}, vol.
  307, no.~20, pp. 1278--1279, 1982, \doi{10.1056/NEJM198211113072025},
  \urlprefix\url{https://doi.org/10.1056/NEJM198211113072025}, pMID: 7133065,
  \eprint{https://doi.org/10.1056/NEJM198211113072025}.

\bibitem{sport-tasks}
F.~Loffing, F.~Sölter, N.~Hagemann, \enquote{Left Preference for Sport Tasks
  Does Not Necessarily Indicate Left-Handedness: Sport-Specific Lateral
  Preferences, Relationship with Handedness and Implications for Laterality
  Research in Behavioural Sciences}, {\em Plos One\/}, vol.~9, no. 105800,
  2014.

\bibitem{streetwar}
S.~SKATEBOARDS, \enquote{Skate stance}, , 2018,
  \urlprefix\url{http://streetwar.org/warfare/tricks/skate-stance/}.

\bibitem{interview}
J.~Marecek, E.~Magnusson, \enquote{Doing Interview-Based Qualitative Research:
  A Learner's Guide}, , 2015.

\bibitem{sample-size2}
J.~Ritchie, L.~Dillon, J.~Lewis, L.~Spencer, \enquote{Quality in Qualitative
  Evaluation: A framework for assessing research evidence}, {\em AdvanceHE\/},
  2003,
  \urlprefix\url{http://www.hollywoodreporter.com/news/earthquake-twitter-users-learned-tremors-226481}.

\bibitem{sample-size3}
M.~Sandelowski, \enquote{Sample size in qualitative research}, {\em Res Nurs
  Health\/}, vol.~18, pp. 179--183, 1995.

\bibitem{sample-size1}
K.~Vasileiou, J.~Barnett, S.~Thorpe, , T.~Young, \enquote{Characterising and
  justifying sample size sufficiency in interview-based studies: systematic
  analysis of qualitative health research over a 15-year period}, {\em BMC\/},
  2018.

\bibitem{wiki}
Wikipedia, \enquote{Comma-separated values}, , 2022,
  \urlprefix\url{https://en.wikipedia.org/wiki/Comma-separated_values}.

\bibitem{chi2}
C.~Light, \enquote{Tutorial: Pearson's Chi-square Test for Independence}, ,
  2008, \urlprefix\url{https://www.ling.upenn.edu/~clight/chisquared.htm}.

\bibitem{review8}
V.~Bewick, L.~Cheek, J.~Ball, \enquote{Statistics review 8: Qualitative data -
  tests of association}, {\em Critical Care\/}, vol.~8, no.~1, 2004.

\bibitem{bland}
M.~Bland, \enquote{An Introduction to Medical Statistics}, {\em
  Physiotherapy\/}, vol.~96, 2010.

\bibitem{review2}
E.~Whitley, J.~Ball, \enquote{Statistics review 2: Samples and populations},
  {\em Critical Care\/}, vol.~6, no. 143, 2002.

\bibitem{IBM}
IBM, \enquote{Cramer's V}, {\em 1\/}, 2021,
  \urlprefix\url{https://www.ibm.com/docs/en/cognos-analytics/11.1.0?topic=terms-cramrs-v}.

\bibitem{odds}
wtskills, \enquote{Odds and Probability || Formula for calculating odds}, ,
  2022, \urlprefix\url{https://wtskills.com/odds-and-probability/}.

\bibitem{oddsratio}
G.~Rodrigues, \enquote{Razão de chances (odds ratio): O que é e o que
  significa?}, {\em PsicoData\/}, 2021,
  \urlprefix\url{https://medium.com/psicodata/razao-de-chances-odds-ratio-o-que-e-e-o-que-significa-8d715961f97d}.

\end{thebibliography}

\balance

\end{document}